\documentclass[11pt]{article}
\usepackage[utf8]{inputenc}
\usepackage{geometry}
\usepackage{amssymb}
\usepackage{upgreek}
\usepackage{setspace}
\usepackage{amsmath}
\usepackage{makecell}
\usepackage{multirow}
\usepackage{adjustbox}
\usepackage{url}
\usepackage{algorithm}
\usepackage{algpseudocode}
\newcommand{\mb}{\mathbf}
\newcommand{\bs}{\boldsymbol}

\DeclareMathOperator*{\argmin}{arg\,min}

\DeclareMathOperator{\EX}{\mathbb{E}}
\usepackage{verbatim}
\usepackage{caption}

\usepackage[compact]{titlesec}
\titlespacing{\section}{0pt}{2ex}{1ex}
\titlespacing{\subsection}{0pt}{1ex}{0ex}
\titlespacing{\subsubsection}{0pt}{0.5ex}{0ex}

\usepackage{natbib}
\usepackage{graphicx}

\usepackage{xr}
\makeatletter
\newcommand*{\addFileDependency}[1]{
  \typeout{(#1)}
  \@addtofilelist{#1}
  \IfFileExists{#1}{}{\typeout{No file #1.}}
}
\makeatother

\begin{document}

\thispagestyle{empty} \baselineskip=28pt \vskip 5mm
\begin{center}
{\Huge{\bf Calibrated Forecasts of Quasi-Periodic Climate Processes with Deep Echo State Networks and Penalized Quantile Regression}}
\end{center}

\baselineskip=12pt \vskip 10mm

\begin{center}\large
Matthew Bonas\footnote[1]{
\baselineskip=11pt Department of Applied and Computational Mathematics and Statistics,
University of Notre Dame, Notre Dame, IN 46556, USA.}, Christopher K. Wikle\footnote[2]{Department of Statistics, University of Missouri, Columbia, MO 65211, USA.} and Stefano Castruccio\textsuperscript{1}
\end{center}

\baselineskip=17pt \vskip 10mm \centerline{\today} \vskip 15mm

\begin{center}
{\large{\bf Abstract}}
\end{center}

\doublespacing

Among the most relevant processes in the Earth system for human habitability are quasi-periodic, ocean-driven multi-year events whose dynamics are currently incompletely characterized by physical models, and hence poorly predictable. This work aims at showing how 1) data-driven, stochastic machine learning approaches provide an affordable yet flexible means to forecast these processes; 2) the associated uncertainty can be properly calibrated with fast ensemble-based approaches. While the methodology introduced and discussed in this work pertains to synoptic scale events, the principle of augmenting incomplete or highly sensitive physical systems with data-driven models to improve predictability is far more general and can be extended to environmental problems of any scale in time or space.

\newpage

\section{Introduction}\label{sec:Intro}

The Earth system comprises of a large number of physical processes through different domains, spatial and temporal scales, and while some of them can be attributed to well known recurring events such as the Earth's revolution and rotation, other variations are more irregular and scale from a few years to multiple decades. While quasi-periodic processes such as the El-Ni$\Tilde{\text{n}}$o Southern Oscillation (ENSO, \cite{phil90}) and Pacific Decadal Oscillation (PDO, \cite{man02}) are important synoptic scale events whose dynamics impact large parts of the world economy and society, their physical understanding, as well as their predictability, are incomplete at best and have been subject of a large number of studies across the past decades in the environmental science community.

Between these aforementioned events, ENSO is the most frequent, with a complete cycle completed every 3-5 years, and will be the main focus of this work (PDO has a considerably longer cycle and is mostly discussed in the supplementary material Section S4). ENSO consists of a warming event (El-Ni$\Tilde{\text{n}}$o) followed by a cooling event (La-Ni$\Tilde{\text{n}}$a) of the sea surface temperature of the equatorial Pacific (see Figure \ref{fig:MERRA2ENSO}) and is considered one of the largest sources of variability in the Earth's climate \citep{vimont05}. Satellite measurements have allowed a comprehensive monitoring of ocean temperatures over the past few decades, and ENSO has been observed to alternate between El-Ni$\Tilde{\text{n}}$o and La-Ni$\Tilde{\text{n}}$a episodes with an irregular cycle. ENSO has far-reaching effects on many countries, including precipitation suppression, warm climate, widespread droughts \citep{cai20} and wildfires which degrade air quality and exacerbates mortality from cardio-respiratory diseases \citep{cri16,mea18}. 

ENSO and other multidecadal events are among the less understood processes of the Earth system, and the degree to which a climate model is able to reproduce them is generally used as a fundamental metric to validate simulations \citep{misra07}. This imperfect knowledge is generally attributable to: 
\begin{enumerate}
    \item a highly nonlinear and parametrization-sensitive physics, at least partly attributable to deep convection of the oceans;
    \item a smaller availability of observational data from ships of opportunity before satellites became available forty years ago, resulting in a smaller confidence in data in the past when validating models dictating multidecadal processes. 
\end{enumerate}

The lack of a good physical representation highlighted in point 1) crucially leads to suboptimal predictive skills from climate models, and calls for the development of data-driven approaches to enhance the learning of the dynamical features and augment physics-informed approaches, as well as for stochastic methods to assess the forecasting uncertainty. Given the complex processes involved in the quasi-oscillation of ENSO, approaches assuming simple linear dynamics have been shown to be comparatively inadequate \citep{knaff97, mcd17}, and more sophisticated nonlinear temporal models are necessary. In fact, over the past few decades there has been a considerable advancement in the literature regarding nonlinear approaches to modelling these complex data (see \cite{pen93, pen98, berliner00, alex08, wikle10, wikle11, patil13, glad14, mcd17, mcd19a, wikle19, petrov20}). Prevalent themes among these works is the use of basis decomposition (through principal components or empirical orthogonal functions), parameter reduction/regularization and the importance of nonlinear spatio-temporal interactions in the models. Parameter estimation for these more sophisticated models is however computationally expensive, especially if one is interested in uncertainty quantification rather than just point estimates. 

In this work, we propose an approach to model ENSO with a new class of calibrated stochastic recurrent Neural Networks (NNs). NNs can naturally capture nonlinear dynamics, but their large parameter space requires considerable computational effort to find even an approximate solution. This is especially problematic in dynamical systems, where timely predictions are critical and computational cost needs to controlled. Instead of relying on a classic, fully non-parametric recurrent NNs, we instead propose a stochastic approximation which assumes that the network is not fixed and unknown, but is rather a sparse realization from a probability distribution. These \textit{Echo State Networks} (ESN, \cite{jae01}) assume that the weight matrix is random with a spike-and-slab distribution for each entry, and hence has two distinctive computational advantages compared to standard recurrent NNs: a) sparse linear algebra can be applied; and b) the parametric space is dramatically reduced. The use of ESNs is effective in reducing the complexity of recurrent NN, but results in additional uncertainty, which needs to be quantified. In previous literature the uncertainty surrounding the ESN forecasts has been assessed using bootstrap/ensemble methods \citep{mcd17} and Bayesian methods \citep{mcd19a, mcd19b}. Additionally, previously work from \cite{bonas21} addressed the issue of uncertainty quantification for the ESN through a calibration approach using monotonic splines. This approach relied on the Gaussianity assumption and aimed properly quantifying the uncertainty of the forecasts whilst also incorporating an expansion of uncertainty through time. This work proposes a more general, new non-parametric approach to calibrate long-range forecasts of the ESN via penalized quantile regression, which allows for a robust method for calibrated forecasting against the assumption of Gaussianity. While the proposed method provides superior predictive skills and calibrated forecast for our applications with ENSO and PDO, it can be applied to any multivariate (possibly spatial) vector whose dynamics are expected to be complex and nonlinear.

\begin{sloppypar}
The rest of the manuscript proceeds as follows: Section \ref{sec:data} introduces the datasets used throughout this work. Section \ref{sec:methods} describes the model and how to perform inference. In Section \ref{sec:Sim} we perform a simulation study with climate model output. In Section \ref{sec:Appl} we show the forecasting and calibration results with observational data and compare it directly to other forecasting methods and the uncertainty quantification methods presented in \cite{mcd17} and \cite{bonas21}. Section \ref{sec:disc} concludes with a discussion on the role of stochastic methods in AI for climate applications, and most importantly the tension between physics informed and data-driven approaches in geosciences. 
\end{sloppypar}

\section{Data}\label{sec:data}

For observations, we consider the Modern-Era Retrospective analysis for Research and Applications version 2 (MERRA2, \cite{gelaro17}), a \textit{reanalysis} data product which assimilates observational data sources with numerical weather models. MERRA2 is comprised of observations from the NASA's Global Modeling and Assimilation Office continuously sampled from 1981\footnote{some observations in late 1980 are also available, but are not considered due to data quality issues}, on a latitude-longitude grid with resolution $0.625^{\circ} \times 0.5^{\circ}$. We also consider simulations from the Coupled Model Intercomparison Project Phase 6 (CMIP6, \cite{IPCC21,eyring16}), and in particular from the scenarioMIP \citep{nei16}, a collection of model simulations providing end-of-century projections. The simulations are performed under historical conditions from 1850 to 2014, and from 2015 to 2100 they follow the Representative Concentration Pathway 2.6 (RCP2.6, \cite{vuu11}), a scenario assuming a low radiative forcing of 2.6 $W\cdot m^{-2}$ by 2100, coupled with the shared socioeconomic pathway scenario 1 (SSP1, \cite{eyring16}), which assumes an adaption of sustainable worldwide emission policies. The choice of RCP2.6-SSP1 was dictated by the large amount of data available, being one of the core simulations in the CMIP6. We choose the Model for Interdisciplinary Research on Climate version 6 (MIROC, \cite{MIROC6}), developed by the Japan Agency for Marine-Earth Science and Technology, Atmosphere and Ocean Research Institute, the National Institute for Environmental Studies, and the RIKEN Center for Computational Science, for which we used thirty ensemble members, each obtained by perturbing atmospheric conditions in 1850, the first simulation year. This ensemble is resolved on a latitude-longitude grid with resolution $1^{\circ} \times 0.5^{\circ}$. For both CMIP6-MIROC and MERRA2, we focus on monthly sea-surface temperature (SST) for the areas of interest to assess the temporal behavior of ENSO and PDO.

\begin{figure}[!bt]
\centering
\includegraphics[height = 10.5cm]{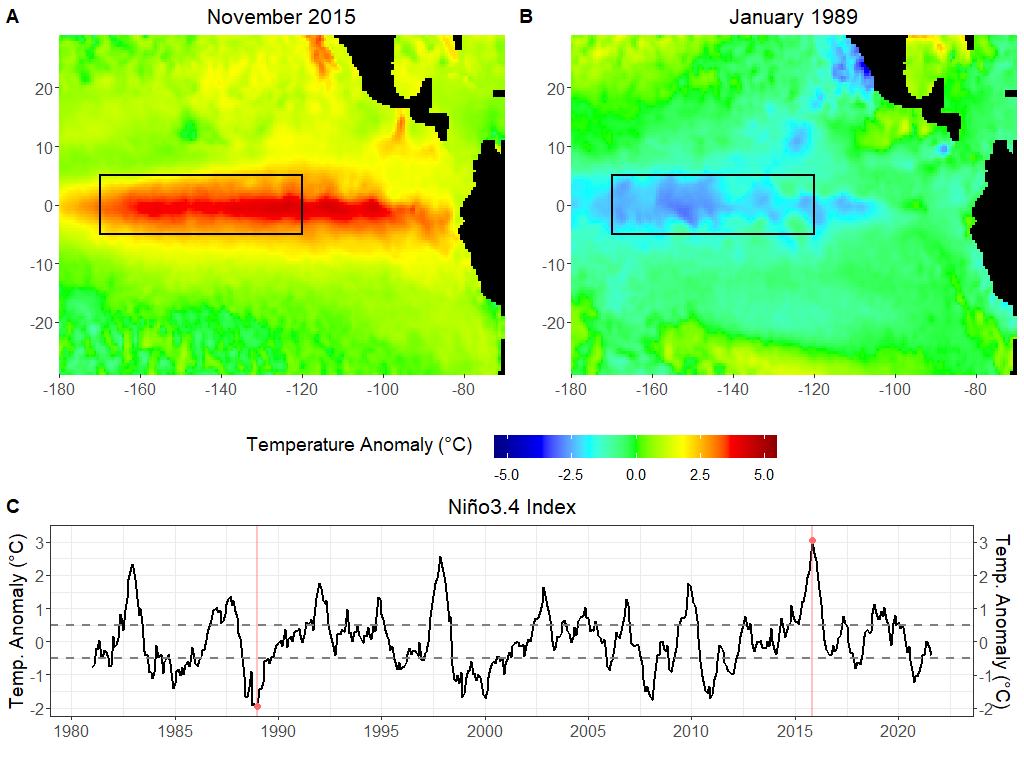}
\caption{Sea Surface Temperature anomaly in the tropical Pacific Ocean during a typical year of (A) El-Ni$\Tilde{\text{n}}$o and (B) La-Ni$\Tilde{\text{n}}$a. Panel (C) shows the entire time series of the Ni$\Tilde{\text{n}}$o-3.4 index from 1981 through August, 2021. The two pink lines represent the time points for the maps shown in (A) and (B), and the gray dashed lines represent the threshold of $\pm0.5^{\circ}\text{C}$, where the onset of the phenomena is declared \citep{noaa}. The black box in panels (A) and (B) represents the Ni$\Tilde{\text{n}}$o-3.4 index region.}
\label{fig:MERRA2ENSO}
\end{figure}

We are interested in assessing the dynamics of ENSO, a quasi-oscillatory process comprised of an anomalous warming (El-Ni$\Tilde{\text{n}}$o) and cooling (La-Ni$\Tilde{\text{n}}$a) of the tropical Pacific Ocean. The Ni$\Tilde{\text{n}}$o-3.4 index is one of the most common indices measuring ENSO, and it is defined as the average SST anomaly in the region between $5^{\circ}$S-$5^{\circ}$N in latitude and $170^{\circ}$W-$120^{\circ}$W in longitude, where a 3-month period above or below a threshold of  $\pm0.5^{\circ}\text{C}$ is when the onset of El-Ni$\Tilde{\text{n}}$o or La-Ni$\Tilde{\text{n}}$a is declared \citep{noaa}. The anomaly is calculated by subtracting the historical average temperature from the current observed temperature to check whether it is above or below the aforementioned threshold. While we are mostly interested in the Ni$\Tilde{\text{n}}$o-3.4 index region (indicated in the black window in Figure \ref{fig:MERRA2ENSO}(A-B)), we consider data on a larger spatial domain ranging from $29^{\circ}$S-$29^{\circ}$N in latitude and $180^{\circ}$W-$70^{\circ}$W in longitude. Besides ENSO, there are other climate indices with a quasi-oscillatory pattern with less frequent occurrences dictated by atmospheric and ocean circulation patterns. Among them, we consider the PDO, which is composed of a recurring pattern of warm and cool SST anomalies in the north Pacific ocean (latitudes above $20^{\circ}$N, \cite{man02}) every 20-30 years, and is shown in Figure S3 (forecasting indices with even longer cycles such as the Atlantic Multi-Decadal Oscillation \citep{keen15} will also be discussed in the supplementary material Section S5). The climate index associated with PDO is the first empirical orthogonal function (EOF) (or principal component analysis in statistical terminology) of the SST anomalies after the global average SST trend has been removed. Here we present the result for ENSO; forecasts and analysis on the PDO are deferred to the supplementary material Section S4.

The spatial and temporal structure of MERRA2 El-Ni$\Tilde{\text{n}}$o and La-Ni$\Tilde{\text{n}}$a can be observed in Figure \ref{fig:MERRA2ENSO}. El-Ni$\Tilde{\text{n}}$o is characterized by a widespread anomalous warming of the SST in the Equatorial Pacific, as apparent in panel (A), while La-Ni$\Tilde{\text{n}}$a is characterized by an anomalous cooling, see panel (B). The quasi-periodic nature of this process can be observed in the time series of the Ni$\Tilde{\text{n}}$o-3.4 index in panel (C), where both the occurrence and magnitude for the two phases are shown. Compared to MERRA2, the ENSO index from the CMIP6-MIROC simulations has a considerably more regular structure, as can be seen in Figure S1 (B) in the supplementary material Section S1 for one ensemble member. Though El-Ni$\Tilde{\text{n}}$o and La-Ni$\Tilde{\text{n}}$a occur in a relatively regular pattern within each individual simulation, among them there is high variability, as shown in Figure S1 (C). 

Throughout this work, we denote by $\mb{Y}_t=(Y_t(1),\ldots, Y_t(N))^\top$ the monthly SST anomaly, where time $t=1,\ldots, T$ indicates the months since the first observation, and $N$ is the total number of vector elements (the spatial locations in this case), which depends on the index considered. When no confusion arises, we will proceed under the assumption that this may indicate either MERRA2 or CMIP6-MIROC. In the simulation study, instead of generating data from a nonlinear dynamical equation, we focus on the ENSO of climate model simulations from the CMIP6-MIROC, since they more closely resembles the real dynamics of MERRA2. Additionally, even though data is available up to today, for MERRA2 we consider $T=456$ (months from January 1981 to December 2018) since this is the most recent time which there was a clear and sharp transition between the phases of ENSO. For CMIP6-MIROC we have $T=3012$ (months from January 1850 to December 2100).  

\section{Methodology}\label{sec:methods}

We introduce the general framework for modelling dynamic data and the recurrent NN model in Section \ref{sec:Modelling}, and we show how to perform inference (learning) in Section \ref{sec:forcinfer}. Finally, we then detail a new approach to calibrate the forecast uncertainty by means of a penalized quantile regression in Section \ref{sec:Calibration}.  

\subsection{Modelling Dynamic Data}\label{sec:Modelling}

Our ultimate goal is to forecast the occurrence of ENSO and other multi-decadal events given their past history. More formally, we aim at determining $\mb{Y}_{t+K}$ for $K>0$ (and the distribution thereof), conditional on $\mb{Y}_{t}, \mb{Y}_{t-1}, \ldots \mb{Y}_{1}$ (there could also be covariates but for the sake of simplicity in the notation we omit them). In a very general sense, the dynamics of the vector $\mb{Y}_{t+1}$ can be expressed as some (possibly stochastic) function of the past history and some input
\begin{equation}\label{eq:dyn}
\mb{Y}_{t+1}=g\left(\mb{Y}_{t},\ldots, \mb{Y}_{1} \mid \bs{\theta}\right),
\end{equation}
for some parameters $\bs{\theta}$, so that $\mb{Y}_{t+K}$ can be obtained iteratively. We assume the $\mb{Y}_{t}$ has no trend since the climate indices we consider rely on anomalies, however the methods presented can be easily generalized to a scenario in which a mean is present. A linear stochastic function $g$ limited to a fixed time horizon represents an interpretable, fast and convenient choice and will be used as our term of comparison throughout this work. If we denote with $Y_t$ the generic element of $\mb{Y}_t$, then for every vector element we can define an auto-regressive moving average model (ARMA$(p,q)$; \cite{box16}) as:
\begin{eqnarray}\label{eqn:ARFIMA}
g(Y_{t}, \dots, Y_{1}) = \sum_{i=1}^{p}\phi^{(AR)}_{i}Y_{t-i} + \sum_{i = 1}^{q}\phi^{(MA)}_{i}\epsilon_{t-i},
\end{eqnarray}

\begin{sloppypar}
\noindent where $p, q$ are positive integers and $\epsilon_{t}\stackrel{i.i.d.}{\sim}N(0,\sigma^2)$, i.e., independent and identically distributed zero mean Gaussian random variables, so that $\bs{\theta}=\left(\left\{\phi^{(AR)}_{i}, i=1, \ldots, p\right\}, \left\{\phi^{(MA)}_{i}, i=1,\ldots, q\right\}, \sigma^2\right)$. This model can be generalized to allow differencing data to capture non-stationary behavior (ARIMA, \cite{brockwell16}) or even more generally fractional differencing to capture some extent of long-range dependence (ARFIMA, \cite{granger80, hosking81}). In this work we also consider a vector autoregressive model (VAR, \cite{hyn21}) for all vector elements simultaneously:
\end{sloppypar}
\begin{equation}\label{eq:var}
 \bs{Y}_{t}=\sum_{i=1}^p \mb{\Phi}_i^{\text{(AR)}}\bs{Y}_{t-i}+\bs{\varepsilon}_{t-i},
\end{equation}
\noindent where now $\mb{\Phi}_i^{\text{(AR)}}$ is an unknown matrix, and the vector error term is such that $\bs{\epsilon}_{t}\stackrel{i.i.d.}{\sim}N(0,\bs{\Sigma})$.

In the case of complex synoptic patterns such as ENSO, the dynamics are far from linear, and while VAR could capture some of this structure through the interaction of the coefficients associated with the primary spatial modes, a full nonlinear model for $g$ could result in a higher degree of flexibility and better predictive performances, as will be shown in Sections \ref{sec:Sim} and \ref{sec:Appl}. 

\subsubsection{Recurrent Neural Networks}\label{sec:RNN}

\quad In machine learning, a general and flexible class of models for $g$ has been proposed by adapting NNs to a dynamic setting and using a hidden state-space to control the temporal behavior, in a similar fashion to state-space models in the statistical literature \citep{kolen01}. These recurrent NNs assume that the function $g$ is described by a set of hidden nodes whose dynamics are nonlinear and controlled by a NN. A popular subclass of recurrent NNs modelling long-range dependence is specified as follows \citep{luk09}:
\begin{subequations}\label{eqn:RNN}
\begin{flalign}
\text{response:} & \qquad \bs{Y}_{t} = f\left(\mb{B}\bs{h}_{t}\right)
\label{eqn:RNN1}, \\
\text{hidden state:} & \qquad \bs{h}_{t} = (1-\alpha)\bs{h}_{t-1} + \alpha\bs{\omega}_{t} \label{eqn:RNN2},\\
& \qquad \bs{\omega}_{t} = f\left(\bs{W} \bs{h}_{t-1} + \bs{W}^{\text{in}}\bs{x}_{t}\right) \label{eqn:RNN3},
\end{flalign}
\end{subequations}

\noindent the data $\bs{Y}_{t}$ is represented in \eqref{eqn:RNN1} through a nonlinear function $f$ and a low rank decomposition of the hidden states. Formally, $\mb{B}$ is a matrix of unknown coefficients, and $\bs{h}_{t}$ is the $n_{h}$-dimensional hidden state vector controlling the dynamics. Part of the model's memory is controlled by the leaking rate parameter $\alpha \in [0,1]$ in \eqref{eqn:RNN2}. The term $\bs{\omega}_{t}$ in \eqref{eqn:RNN3} is computed as the function $f$ applied to the combination of the past state $\bs{h}_{t-1}$ and the $n_{x}$-dimensional covariate vector, which in this work consists of past observations up to $m$ time steps: $\bs{x}_{t}=(\bs{1},\bs{Y}^{\top}_{t-1},\ldots,\bs{Y}^{\top}_{t-m})^{\top}$. The \textit{activation} function $f$ is applied element-wise and has a pre-specified form such as hyperbolic tangent or rectified linear unit \citep{goo16}. 

The flexibility of these recurrent NNs in capturing the dynamics of $\mb{Y}_t$ comes at the expense of a sizable computational cost for inference due to the large dimension of the parameter vector $\bs{\theta}=\left(\mb{B}, \bs{W}, \bs{W}^{\text{in}}, \alpha \right)$. Inference is further complicated by the emergence of vanishing (or exploding) gradients in gradient-based minimization algorithms (\textit{backpropagation}) stemming from the recursive nature of the model \citep{beng94}. Additionally, while a point forecast is operationally feasible, the quantification of its uncertainty requires the evaluation of the uncertainty of the parameter estimates. Methods have been developed for standard NNs such as dropout or dilution \citep{hin12}, or Bayesian networks \citep{blu15, gra11}, but while in a static setting the associated additional computational burden may be acceptable, in a dynamic setting timely inference is fundamental, as results are valuable only to the extent that they can be provided at a fraction of the sampling rate. 

\subsubsection{Deep Echo-State Networks}\label{sec:ESN}

In order to retain the flexibility of recurrent NNs while avoiding the high computational cost implied by large unknown matrices, previous work has proposed a faster, stochastic version aimed at reducing the dimensionality of the parameter space. These ESNs assume that both $\bs{W}$ and $\bs{W}^{\text{in}}$ are random with a distribution controlled by some parameters, thereby dramatically reducing the model dimensionality, while at the same time avoiding the instabilities associated with computing gradients from backpropagation, as a small parametric space can be estimated with a cross-validation approach. In fact, in a series of recent works \citep{mcd17, mcd19a, mcd19b} the ESN was introduced to the statistical community for modeling spatio-temporal processes, generating long-lead SST forecasts, and for uncertainty quantification. The first of these works introduced a one layer (shallow) ESN whereas the latter works advanced these models through Bayesian uncertainty quantification as well as a deep ESN (DESN). A DESN can be formulated as follows \citep{mcd19b}:
\begin{subequations}\label{eqn:DESN}
\begin{flalign}
\text{output:} & \qquad \mb{Y}_{t} = \mb{B}_{D}\mb{h}_{t, D} + \sum_{d=1}^{D-1} \mb{B}_{d}k\left(\Tilde{\mb{h}}_{t,d}\right) + \bs{\epsilon}_{t} \label{eqn:DESN1}, \\
\text{hidden state $d$:} & \qquad \mb{h}_{t,d} = (1-\alpha)\mb{h}_{t-1,d} + \alpha\bs{\omega}_{t,d} \label{eqn:DESN2},\\
& \qquad \bs{\omega}_{t,d} = f_{h}\left(\frac{\nu_{d}}{|\lambda_{W_{d}}|} \mb{W}_{d} \mb{h}_{t-1,d} + \mb{W}_{d}^{\text{in}}\Tilde{\mb{h}}_{t,d-1}\right), \text{ for } d > 1 \label{eqn:DESN3},\\
\text{reduction $d-1$:} & \qquad \Tilde{\mb{h}}_{t,d-1} \equiv Q\left(\mb{h}_{t,d-1}\right), \text{ for } d > 1 \label{eqn:DESN4},\\
\text{input:} & \qquad \bs{\omega}_{t,1} = f_{h}\left(\frac{\nu_{1}}{|\lambda_{W_{1}}|} \mb{W}_{1} \mb{h}_{t-1, 1} + \mb{W}_{1}^{\text{in}}\mb{x}_{t}\right) \label{eqn:DESN5},\\
\text{matrix distribution:} & \qquad W_{d_{i,j}} = \gamma^{W_{d}}_{i,j} p(\eta_{W_{d}}) + (1-\gamma^{W_{d}}_{i,j})\delta_{0} \label{eqn:DESN6},\\
& \qquad W^{\text{in}}_{d_{i,j}} = \gamma^{W_{d}^{\text{in}}}_{i,j} p(\eta_{W_{d}^{\text{in}}}) + (1-\gamma^{W_{d}^{\text{in}}}_{i,j})\delta_{0} \label{eqn:DESN7},\\
& \qquad \gamma^{W_{d}}_{i,j} \text{\;$\thicksim$\;} Bern(\pi_{W_{d}}), \quad \gamma^{W_{d}^{\text{in}}}_{i,j} \text{\;$\thicksim$\;} Bern(\pi_{W_{d}^{\text{in}}}), \nonumber
\end{flalign}
\end{subequations}

\noindent which comprises of $d = \{1, \dots, D\}$ layers (depth). The weight matrices $\bs{W}_{d}$ and $\bs{W}^{\text{in}}_{d}$ in \eqref{eqn:DESN6} and \eqref{eqn:DESN7} are randomly generated and sparse, with an independent entry-wise spike-and-slab prior, i.e., with zero value with a given probability ($\pi_{W_{d}}$ and $\pi_{W_{d}^{\text{in}}}$ for $\bs{W}_{d}$ and $\bs{W}^{\text{in}}_{d}$, respectively), otherwise drawn from a symmetric distribution $p(\cdot)$ centered about zero \citep{luk12}. For this work, it is assumed that $p(\cdot)$ is a standard normal distribution, thus avoiding the need to specify parameters $\eta_{W_{d}}$ and $\eta_{W_{d}^{\text{in}}}$, but other choices such as the uniform distribution have been proposed \citep{mcd17, huang2021}. The DESN must also respect the \textit{echo-state property}: after a sufficiently long time sequence, the model must asymptotically lose its dependence on the initial conditions  \citep{luk12,jae07}. This property holds when the largest eigenvalue of $\bs{W}_{d}$, known as its spectral radius and denoted by $\lambda_{W_{d}}$, is less than one. To establish this property, the scaling parameter $\nu_{d}$ is introduced to ensure that the spectral radius does not violate this condition on equation \eqref{eqn:DESN3}. Additionally, the error $\bs{\epsilon}_{t}$ in \eqref{eqn:DESN1} is assumed to be independent identically distributed in time and is expressed as a mean zero multivariate normal. The function $Q(\cdot)$ in equation \eqref{eqn:DESN4} represents a dimension reduction function of the hidden state $\mb{h}_{t,d}$ resulting in a $n_{\Tilde{h}, d}$ dimension reduced hidden state $\mb{\Tilde{h}}_{t,d}$. For this work, we use an empirical orthogonal function (EOF) approach to act as $Q(\cdot)$ in the dimension reduction stage. \cite{mcd17} showed how including quadratic interactions in the output term \eqref{eqn:DESN1} can improve the forecast accuracy of the model. However, it was found in \cite{bonas21} that these quadratic interactions may not be necessary in all applications, hence we only show the linear case for simplicity here. Thus, the output $\mb{Y}_{t}$ is expressed as a linear function $\mb{B}_{D}\mb{h}_{t, D} + \sum_{d=1}^{D-1} \mb{B}_{d}k_{h}\left(\Tilde{\mb{h}}_{t,d}\right) + \bs{\epsilon}_{t}$ in \eqref{eqn:DESN1}, where $\mb{h}_{t,D}$ is an $n_{h,D}$-dimensional state vector, $\mb{\Tilde{h}}_{t,d}$ is an $n_{\Tilde{h},d}$-dimensional state vector, and the $\mb{B}_{d}$ are unknown coefficient matrices which are estimated via ridge regression with penalty $\lambda_{r}$. The $k(\cdot)$ in equation \eqref{eqn:DESN1} represents an activation function to ensure that the values of $\Tilde{\mb{h}}_{t,d}$ are on a similar scale to those within $\mb{h}_{t,D}$ (i.e., the range of values which the two vectors take is similar or the same). With an abuse of notation, we denote the model hyper-parameters by $\bs{\theta}=(n_{h,d}, n_{\Tilde{h}, d}, m, \nu_{d},\lambda_{r}, \pi_{W_{d}}, \pi_{W_{d}^{\text{in}}}, \alpha, D)$, excluding $\mb{B}_d$, which can be estimated separately as will be shown in the next section. 

\subsection{Inference and Forecasting}\label{sec:forcinfer}

In order to perform inference on the parameter vector $\bs{\theta}$, a portion of the observed training data is held out as a validation set and cross-validation is performed. For a fixed $\bs{\theta}$, the matrices $\mb{B}_{d}$, for $d=1, \dots,D$, from equation \eqref{eqn:DESN1} are estimated and predictions are produced for the validation set, the mean squared error (MSE) is calculated and the optimal value for $\bs{\theta}$ is determined by the set of hyper-parameters in $\bs{\theta}$ which minimizes it. We estimate the coefficient matrices $\mb{B}_{d}$ through ridge regression with penalty parameter $\lambda_{r}$. More specifically, we denote equation \eqref{eqn:DESN1} in the matrix form of $\textbf{Y} = \textbf{HB} + \textbf{E}$, where
\begin{gather}
 \textbf{Y} = \begin{bmatrix} \textit{\textbf{Y}}^{\top}_{1} \\ \vdots \\ \textit{\textbf{Y}}^{\top}_{T} \end{bmatrix}, \quad
 \textbf{H} = \begin{bmatrix} \textit{\textbf{h}}^{\top}_{1} \\ \vdots \\  \textit{\textbf{h}}^{\top}_{T} \end{bmatrix},  \quad
 \textbf{E} = \begin{bmatrix} \mbox{\boldmath{$\epsilon$}}^{\top}_{1} \\ \vdots \\ \mbox{\boldmath{$\epsilon$}}^{\top}_{T}. \end{bmatrix}.
 \nonumber
\end{gather}

\noindent Here $\mb{h}^{\top}_{t} = \left(\mb{h}_{t,D}, k(\mb{\Tilde{h}}_{t,1}), \dots,  k(\mb{\Tilde{h}}_{t,D-1})\right)^{\top}$, for $t=1,\dots,T$ and $\mb{B} = \left(\mb{B}_{D}, \mb{B}_{1}, \dots, \mb{B}_{D-1}\right)^{\top}$. The ridge estimator $\hat{\textbf{B}} = \argmin_{\textbf{B}}(\|\textbf{Y} - \textbf{HB}\|^{2}_{F} + \lambda_{r}\|\textbf{B}\|^{2}_{F})$, where $\|\cdot\|_{F}$ is the Frobenius norm, has a closed form $\hat{\textbf{B}} = (\textbf{H}^{\top}\textbf{H} + \lambda_{r}\textbf{I})^{-1}\textbf{H}^{\top}\textbf{Y}$.

The proposed model could in theory be directly used to produce forecasts, however this would require a vector $\bs{x}_{t}$ containing thousands of spatial locations at each time $t$, and hence a high-dimensional vector of hidden states $\bs{h}_{t}$ to capture the nonlinear dynamics of the data. As a result $\mb{B}_{d}, \bs{W}_{d}$ and $\bs{W}^{\text{in}}_{d}$ would all be high-dimensional matrices, increasing the computational cost of forecasting. We chose to retain 20 EOFs as opposed to the more parsimonious choice of 10 in \cite{mcd17} since we aim at capturing more variation in the original data and the implied computational overhead was still manageable. This dimension reduction approach allows us to identify and capture the main sources of spatial variability, which is especially useful in ENSO and its distinct yet irregular modes \citep{glad14, mcd17}. Therefore, in this analysis the output $\bs{Y}_{t}$ in equation \eqref{eqn:DESN1} will have twenty elements, each representing an EOF coefficient. Once the forecasts are produced, they are converted back to their original scale to recreate the estimated spatial field.

Once the DESN parameters are estimated, an ensemble of forecasts at $T+1$ given observations at times $T, T-1, \ldots, 1$ is produced, each prediction corresponding to a different realization of $\bs{W}_{d}$ and $\bs{W}^{\text{in}}_{d}$. Afterwards the ensemble average is treated as the input $\bs{x}_{T+1}$ to forecasts point $T+2$. This approach is then applied iteratively until forecast $T+K$, where $K$ is the number of future time points we would like to predict. 

\subsection{Calibration of the Forecasts}\label{sec:Calibration}

In this work we calibrate each EOF coefficient separately and then reconstruct the spatial field thus, for simplicity, we drop the vector notation and assume that we forecast a scalar process $Y_{t+1}$ given $\left\{Y_{t}, Y_{t-1},\dots\right\}$ is written as $\hat{Y}^{t}_{t+1} = \EX(Y_{t+1}|Y_{t}, Y_{t-1}, \dots)$, where the expectation is computed with respect to the DESN model in \eqref{eqn:DESN}. The use of DESN allows for fast forecasts with a parsimonious yet flexible model, but comes at the expense of additional uncertainty stemming from the random matrices \eqref{eqn:DESN4}. The error needs to be properly quantified, and in this work we propose a new method for long-range forecasting which relies on three key ideas: 1) the uncertainty generated by the DESN can be assessed by performing ensemble forecasting on the data in the training set; 2) the uncertainty can then be controlled to allow its increase as the forecast horizon increases with an appropriate penalty on quantile regression; and 3) marginal forecasts can be calibrated using a \textit{post hoc} approach to adjust the prediction interval (PI) bands and achieve proper coverage. 

In order to calibrate the forecast, we generate an ensemble where each member is obtained through a different realization of $\bs{W}_{d}$ and $\bs{W}^{\text{in}}_{d}$, for time points $T-n_{w}\times K + 1$ through $T$ across multiple windows $w=1,\ldots, n_w$ in intervals of some specified size $K$. The residuals for each ensemble member are computed as
\begin{equation}\label{eqn:residcalc}
{R}^{T-w\times K}_{t+1} = {Y}_{t+1}-\hat{{Y}}^{T-w\times K}_{t+1},
\end{equation}
\noindent and we indicate with $\mb{R}_w=\left({R}^{T-w\times K}_{t+1}\right)_{t}$ for $t=T-(w\times K),\dots,T-\{(w-1)\times K-1\}$, the $K$-dimensional vector of residuals. Since the proposed approach is performed independently for every $w$ and the final intervals are averaged across the windows, for simplicity we omit this parameter from the notation. The residual uncertainty and its change with lead time is then estimated independently using quantile regression \citep{koen05}, which allows  robustness to some degree of perturbation of Gaussianity in the DESN \eqref{eqn:DESN} (see Figure S2 which shows the Gaussian diagnostics), unlike the method presented in \cite{bonas21} which requires this assumption to hold. The $q$-quantile at lead time $k$ $\mu(k;q)$ must avoid \textit{quantile crossing}, i.e., for all $q_{1} \leq q_{2}$, we must have $\mu(k;q_{1}) \leq \mu(k;q_{2})$. To penalize and ultimately avoid quantile crossing in the inference, we simultaneously estimate $\mu(k;q)$ for every $k$ with an appropriate penalized basis spline, i.e., $\mu(k;q) = \bs{S}\bs{a}_{q}$ where $\bs{S}$ is a basis of B-splines and $\bs{a}_{q}$ is a coefficient vector to the estimated \citep{schn12}. The goal is then to minimize 
\begin{flalign}\label{eqn:quantsheet}
& \left\|\bs{R}-\bs{S}\bs{a}_{q}\right\|^{2} + \lambda\left\|(\Delta^2 \bs{D})\bs{a}_{q}\right\|^{2}
\end{flalign}

\noindent with respect to the coefficient vector $\bs{a}_{q}$. The first term represents the $L^2$ distance between $\mb{R}$ and its basis representation, while the second term is the penalty controlled by $\lambda$, where $\Delta^2 \bs{D}$ is a matrix consisting of the column-wise second order differencing of a diagonal matrix $\bs{D}$. This minimization method is more commonly referred to as \textit{quantile sheets} as it allows for the quantiles to be estimated simultaneously, while avoiding quantile crossing \citep{schn12}.

\begin{sloppypar}
In order to encourage the expansion of the forecast uncertainty in time, i.e., $\left|{\hat{\mu}(k, q_{1})} - {\hat{\mu}(k,q_{2})}\right|$ increasing in $k$, instead of assuming that $\bs{D}$ is an identity matrix which uniformly penalizes changes in the (discretized) second order derivative \citep{schn12}, we impose a progressively stricter penalty in $k$. This choice avoids an excessive variation of $\hat{\mu}(k, q)$ for large $k$, thereby leading to an expansion of the distance between quantiles. Specifically, the matrix $\bs{D}$ is assumed to be diagonal with entries from 1 through $s$, the number of columns in the basis matrix $\bs{S}$, scaled by a parameter $\xi>0$. In other words, we have that $D=\text{diag}(d_i)$ with
\begin{equation}\label{eqn:diagpenalty}
d_i=\frac{i}{\xi}, i = \{1, \dots, s\}.
\end{equation}
\end{sloppypar}

While in Section \ref{sec:Appl} we show how the proposed approach empirically allows for expansion of the uncertainty in time, there is no guarantee that the uncertainty indicated by the PIs is reflective of the correct amount of uncertainty from the data. To ensure proper coverage, we propose an additional calibration step. After $\hat{\mu}(k,q)$ is obtained, the distance between each quantile curve and the median $q = 0.5$ is calculated and the interval is obtained by adding (or subtracting) this expression to the forecast median. Once these intervals are generated, an adjustment scalar parameter $\zeta$ is added to the upper and lower interval. This $\zeta$ parameter is optimized via grid search so that proper coverage is achieved on the observed data. We denote the optimal distance between the median and the upper and lower bounds as $\bs{I}_{U}$ and $\bs{I}_{L}$ and define them as:
\begin{flalign}
& \bs{L} =\{L_1, \ldots, L_K\}, L_k=\hat{\mu}(k, 0.5) - \hat{\mu}(k,q_{2}), 0.5 > q_{2}, \implies \bs{I}_{L} = \bs{L} + \zeta,\label{lowerdist}\\
& \bs{U} =\{U_1, \ldots, U_K\}, U_k= {\hat{\mu}(k, q_{1})} - {\hat{\mu}(k,0.5)}, q_{1} > 0.5, \implies \bs{I}_{U} = \bs{U} + \zeta. \label{upperdist}
\end{flalign}

The final interval distance is calculated by averaging $\bs{I}_{L}$ and $\bs{I}_{U}$ over the $n_{w}$ windows. These final distances  ($\bar{\bs{I}}_{L}$ and $\bar{\bs{I}}_{U}$), would then be added (subtracted) from the future forecasts to generate the PIs. In the simulation and application Sections \ref{sec:Sim} and \ref{sec:Appl} it will be shown how the choice of the penalty in \eqref{eqn:quantsheet}, the scaling parameter in \eqref{eqn:diagpenalty}, as well as the interval adjustment will 1) promote an expansion of the uncertainty bands through time; and 2) allow the empirical coverage from PIs to approach the nominal coverage. The adjustment factor $\zeta$ is homogeneous in lead time for simplicity but also because a penalized method with a lead-specific bias with expansion of the uncertainty would be considerably more involved. Given the results obtained, we didn't feel that this generalization it was necessary to implement. For this work we chose $q_1=0.975$ and $q_2=0.025$ in order to focus on a 95\% interval, however any other choice could be made if a different coverage interval is sought. The full approach detailed prior is outlined in Algorithm \ref{alg:UncAdj}.

\begin{algorithm}[!bt]
\caption{Marginal forecast calibration.}
\label{alg:UncAdj}
\textbf{Data:} ${Y}_{t}: t = 1, \dots, T$\\
\textbf{Results:} Upper and lower PI distances, $\bar{\bs{I}}_{U}$ and $\bar{\bs{I}}_{L}$, for forecasts ${\hat{{Y}}}$\\
\textbf{Initialize:} Generate forecasts $\hat{{Y}}^{T-w\times K}_{t+1}$ where $t=T-(w\times K),\dots,T-\{(w-1)\times K-1\}$ and $w=1,\dots,n_{w}$ from the DESN in equation \eqref{eqn:DESN}
\begin{algorithmic}
 \For{$w = 1, \dots, n_{w}$}
     \begin{itemize}
         \item    Calculate the residuals for the forecasts $\mb{R}_w$ from equation \eqref{eqn:residcalc}\;
       \item Fit the residuals using a penalized quantile regression for the desired quantiles outlined in equations \eqref{eqn:quantsheet} and \eqref{eqn:diagpenalty}. Denote the fitted quantiles as $\hat{\mu}(k, q_{1})$ and $\hat{\mu}(k,q_{2})$\;
       \item Calculate the distance between the median $q=0.5$ and each fitted quantile curve $\hat{\mu}(k, q_{1})$ and $\hat{\mu}(k,q_{2})$. Denote these as $\bs{U}$ and $\bs{L}$\;
      \item Calculate $\bs{I}_{L}$ and $\bs{I}_{U}$ from equations \eqref{lowerdist} and \eqref{upperdist} by optimizing $\zeta$ such that there is proper coverage on the observed data.
      \end{itemize}
 \EndFor
\State Calculate $\bar{\bs{I}}_{L} = \frac{1}{n_{w}}\sum_{i=1}^{n_{w}}\bs{I}_{L,i}$ and $\bar{\bs{I}}_{U} = \frac{1}{n_{w}}\sum_{i=1}^{n_{w}}\bs{I}_{U,i}$, the average interval distance across the $n_{w}$ windows.
\end{algorithmic}
\end{algorithm}

\section{Simulation Study: CMIP6-MIROC}\label{sec:Sim}

In this Section we show how the proposed stochastic machine learning approach yields superior predictive performance compared to other reference models, and that the uncertainty stemming from the use of stochastic matrices in DESN can be calibrated compared to other approaches presented like those in \cite{mcd17} and \cite{bonas21} This Section describes the forecasting for the CMIP6-MIROC simulations as well as the uncertainty quantification. In this Section we focus on ENSO, and in the supplementary material Section S4 we show the results for the PDO. 
\begin{figure}[!b]
\centering
\includegraphics[width = 14cm]{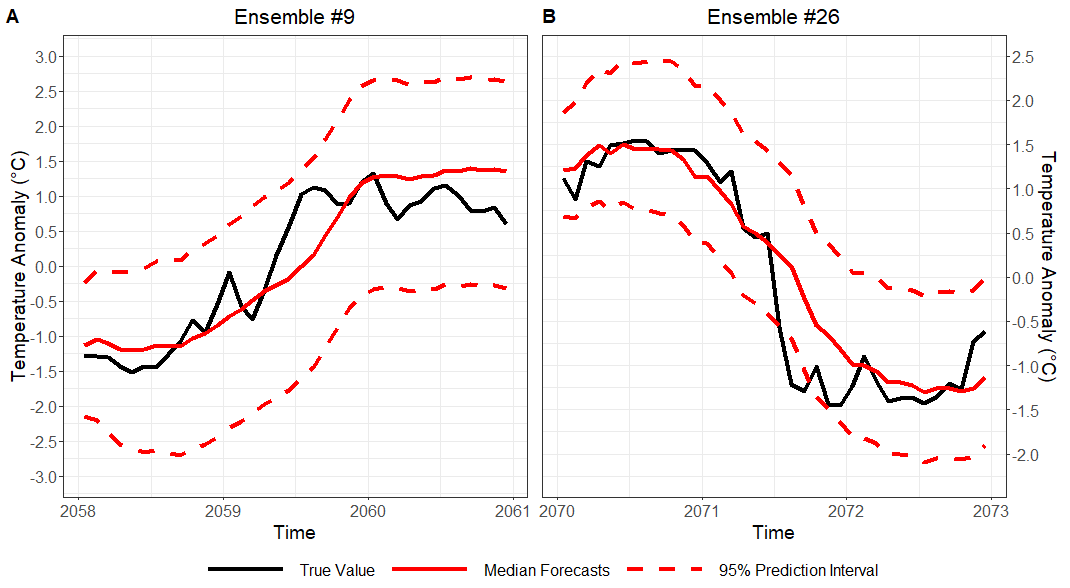}
\caption{Forecasts and uncertainty quantification for the Ni$\tilde{\text{n}}$o 3.4 index of (A) ensemble member 9 and (B) ensemble member 26 from the CMIP6-MIROC data with the DESN and the penalized quantile regression method.}
\label{fig:N34Ensemb}
\end{figure}

We perform forecasting for the simulated CMIP6-MIROC monthly data introduced in Section \ref{sec:data} with the DESN model. For each ensemble member, the date ranges were chosen to represent a clear and sharp transition between the two ENSO phases, as exemplified by the two ensemble members in Figure \ref{fig:N34Ensemb}. The grids used for the hyper-parameter selection for the DESN from Section \ref{sec:ESN} are as follows: $n_{h, d}=$\{$60, 90, 120, 150$\}, $n_{\Tilde{h}}=$\{$15, 20, 25,\dots, 55, 60$\}, $m = $\{$1, \dots, 4$\}, $\nu_{d} = $\{$0.2 \times s_{\nu_{d}} : s_{\nu_{d}} = 1, \dots, 5$\}, $\lambda_{r}$ = \{$0.001,0.01,0.1$\}, $\alpha = $\{$0.01 \times s_{\alpha} : s_{\alpha} = 1, \dots, 100$\}, $\pi_{W} = $\{$0.1 \times s_{\pi_{W}} : s_{\pi_{W}} = 1, \dots, 10$\}, $\pi_{W^{\text{in}}} = $\{$0.1 \times s_{\pi_{W^{\text{in}}}} : s_{\pi_{W^{\text{in}}}} = 1, \dots, 10$\} and we chose $D = \{1,2,3\}$ layers. The DESN model hyper-parameters were optimized for each ensemble member independently using the MSE on a validation set composed of the last $\tilde{K}=36$ points prior to the testing points using the inference approach described in Section \ref{sec:forcinfer}. Additionally, we used $n_{w} = 10$ windows in the training data to optimize the upper and lower PI bands described in the calibration approach in Section \ref{sec:Calibration}.

The forecast median of the Ni$\tilde{\text{n}}$o 3.4 index, along with the calibrated 95\% PIs for two ensemble members of the CMIP6-MIROC data can be seen in Figures \ref{fig:N34Ensemb}A and \ref{fig:N34Ensemb}B. Additionally, the quantile regression method presented in Section \ref{sec:Calibration} drastically improves the uncertainty quantification, as shown in Table \ref{tbl:UQComparison}. Specifically, the 95\% PIs using the ensemble-based method presented in \cite{mcd17} only capture an average of 68.1\% (standard deviation 20.8\%) of the true data across all ensemble members. Using the monotonic spline method proposed in \cite{bonas21}, the 95\% PIs capture an average of 99.1\% (2.4\%) of the true data across the ensemble members, hence substantially over-estimating the uncertainty. After applying the calibration method presented in Section \ref{sec:Calibration}, the corrected 95\% PIs capture an average 95.8\% (7.7\%) of the true data, with an appreciable expansion of the uncertainty through time as apparent in Figure \ref{fig:N34Ensemb}. We also evaluate the DESN performance on forecasting the Ni$\tilde{\text{n}}$o 3.4 index versus an ARFIMA and VAR models using the MSE and the continuous ranked probability score (CRPS; \cite{gneit07}), a scoring rule for probabilistic forecasts which equals zero in the event of a perfect forecast. Additionally, we use a hidden Markov model (HMM, \cite{rab89}), a long short-term memory (LSTM, \cite{hoch97}) model and a gated recurrent unit (GRU, \cite{cho14}) model for additional comparison to the DESN. Descriptions of the HMM, LSTM, and GRU models have been deffered to the supplementary material Section S3 The metrics corresponding to each of these models can be seen in Table \ref{tbl:MethodComp}, where the mean and standard deviation for each metric is reported. From this table it can be seen that the DESN showed improved point predictions with a mean MSE of 0.43 (0.31) and CRPS of 0.39 (0.10) across all ensemble members which outperforms all other reference models. The next best model (that isn't an ESN model) was VAR which returned values of 0.68 (0.22) for MSE and 0.48 (0.08) for CRPS.

\begin{table}[!b]
\begin{centering}
\begin{tabular}{||c|c|c||} 
\hline
Forecasting Method & MSE & CRPS \\ [1ex] 
\hline\hline
Deep ESN & 0.43 (0.31) & 0.68 (0.14)  \\
\hline
Shallow ESN & 0.87 (0.77) & 0.74 (0.21)  \\
\hline
ARFIMA & 0.84 (0.43) & 0.85 (0.19)  \\
\hline
Vector Autoregression & 0.68 (0.22) & 0.81 (0.144) \\ 
\hline
Hidden Markov Model & 1.04 (0.48) & 0.88 (0.19) \\
\hline
Long Short-Term Memory & 1.02 (0.59) & 0.74 (0.25) \\
\hline
Gated Recurrent Unit & 1.01 (0.55) & 0.73 (0.24) \\[1ex]
\hline
\end{tabular}
\caption{Comparison of the forecasting methods in terms of MSE and CRPS for the Ni$\Tilde{\text{n}}$o-3.4 index for all realizations of the CMIP6-MIROC data. A mean across all realizations is shown, and in parenthesis the standard deviation is reported.}
\label{tbl:MethodComp}
\end{centering}
\end{table}

\begin{figure}[!bt]
\centering
\includegraphics[width = 14cm]{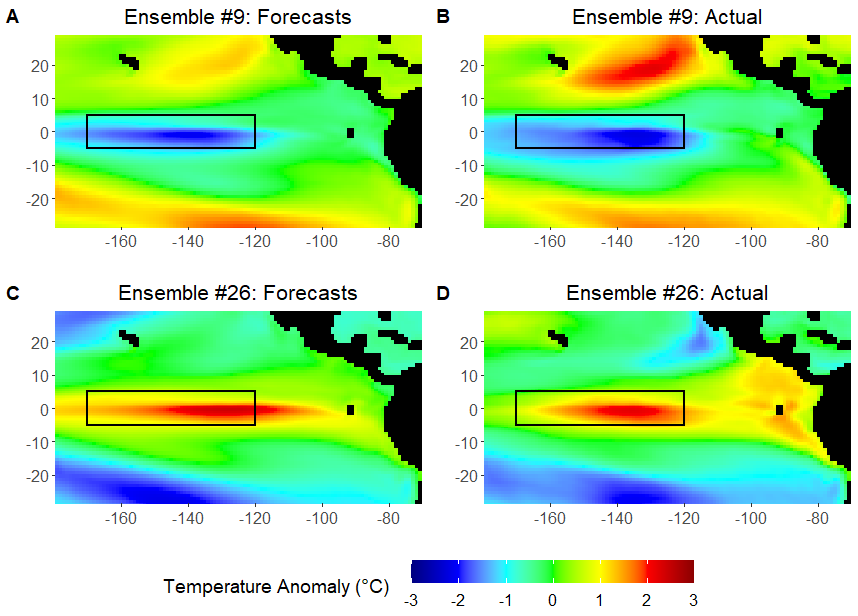}
\caption{For ensemble member 9 of the CMIP6-MIROC data, at a peak La-Ni$\tilde{\text{n}}$a event in January 2058, we show the SST anomaly (A) forecast median and (B) true data. For ensemble member 26 of the CMIP6-MIROC data, at a peak El-Ni$\tilde{\text{n}}$o event in October 2070, we show the SST anomaly (C) forecast median and (D) true data. The black box in all panels represents the Ni$\Tilde{\text{n}}$o-3.4 index region.}
\label{fig:FullRegionForcs}
\end{figure}

While the Ni$\tilde{\text{n}}$o 3.4 index provides relevant aggregate information, it is also of scientific interest to  forecast over the entire ENSO spatial domain. From the spatial grid created from the EOFs, the model forecasts largely capture the ENSO magnitude. It can be seen in Figure \ref{fig:FullRegionForcs} how the DESN is able to accurately predict the SST anomalies across the whole spatial domain for a peak La-Ni$\tilde{\text{n}}$a event (panel A) and a peak El-Ni$\tilde{\text{n}}$o event (panel C) and  for the same ensemble members as before. These figures ultimately show how the proposed approach is able to accurately forecast not only the Ni$\tilde{\text{n}}$o 3.4 index but also across the spatial domain of interest.

\section{Application: MERRA2}\label{sec:Appl}

In this Section we apply the DESN calibration approach from Section \ref{sec:Calibration} to MERRA2 data and compare its predictive performance with currently available models and uncertainty quantification with the methods presented in \cite{mcd17} and \cite{bonas21} We focus on ENSO, and in the supplementary material Section S4 we show the results for the PDO. 

The MERRA2 data have $T=420$ points available (January 1981 through December 2015) as the training set, with the subsequent $K=36$ points (January 2016 through December 2018) used as the testing set. Similarly to the CMIP-MIROC simulated data in the previous section, these times were chosen as they exemplify a clear and fast transition between El-Ni$\Tilde{\text{n}}$o and La-Ni$\Tilde{\text{n}}$a. The parameters were optimized using the same approach for the CMIP6-MIROC data, where the last $\tilde{K}=36$ points in the training set (January 2013 through December 2015) were used as the validation set to optimize the hyper-parameters of the model. The grids used for hyper-parameter selection were the same as those used for the CMIP6-MIROC data. The estimated hyper-parameters from MSE cross-validation are: $\hat{D} = 2$, $\hat{n}_{h}= 90$, $\hat{n}_{\Tilde{h}} = 50$, $\hat{m}=2$, $\hat{\lambda}_{r} = 0.1$, $\hat{\nu} = \{0.20, 0.60\}$, $\hat{\pi}_{W} = 0.40$, $\hat{\pi}_{W^{\text{in}}} = 0.60$ and $\hat{\alpha} = 0.59$. Additionally, we used $n_{w} = 5$ windows in the training data to optimize the upper and lower PI bands described in the calibration approach in Section \ref{sec:Calibration}.

For MERRA2, the training and testing dataset contained up to 8 ENSO cycles. However, the DESN still outperforms each of the other models. Indeed, the MSE and CRPS for the DESN were 0.62 and 0.57, respectively, whereas these metrics for the next two best models, HMM and VAR, were 0.93 and 0.65 for HMM and 0.99 and 0.81 for VAR as can be seen in Table \ref{tbl:MERRA2MethodComp}. Thus, this shows an improvement of at least 33\% in MSE and 12\% in CRPS versus the next best models. The results in Table \ref{tbl:MERRA2MethodComp} also indicate that the shallow ESN (i.e. $D=1$ in equation \eqref{eqn:DESN}) does not perform well on the MERRA2 data. In fact, the shallow ESN performs worse in terms of MSE than the HMM. These results ultimately justify the need for a deeper, more complex model to forecast the MERRA2 data. Hence, the use of the DESN is not only justified by its own forecasting performance, but also by the performance of a shallower, less complex version of the ESN.

\begin{table}[!t]
\begin{centering}
\begin{tabular}{||c|c|c||} 
\hline
Forecasting Method & MSE & CRPS \\ [1ex] 
\hline\hline
Deep ESN & 0.62 & 0.57  \\
\hline
Shallow ESN & 0.92 & 0.59  \\
\hline
ARFIMA & 1.14 & 0.84 \\
\hline
Vector Autoregression & 0.99 & 0.81 \\ 
\hline
Hidden Markov Model & 0.93 & 0.65 \\
\hline
Long Short-Term Memory & 2.50 & 1.39 \\
\hline
Gated Recurrent Unit & 2.46 & 1.38 \\[1ex]
\hline
\end{tabular}
\caption{Comparison of the forecasting methods in terms of MSE and CRPS for the Ni$\Tilde{\text{n}}$o-3.4 index for the MERRA2 data}
\label{tbl:MERRA2MethodComp}
\end{centering}
\end{table}

\begin{table}[!b]
\begin{centering}
\begin{tabular}{||c|c|c||} 
\hline
\multirow{2}{*}{Calibration Approach} & \multicolumn{2}{c||}{Empirical 95\% Coverage} \\ \cline{2-3} &  CMIP6-MIROC & MERRA2\\ [1ex] 
\hline\hline
Ensemble-Based & 68.1 (20.8) & 61.1  \\
\hline
Monotonic Spline & 99.1 (2.4) & 77.8 \\
\hline
Quantile Regression & 95.8 (7.7) & 91.7  \\
\hline
\end{tabular}
\caption{Uncertainty quantification methods compared in terms of their nominal 95\% PI coverage on both the CMIP6-MIROC and MERRA2 data. The ensemble-based approach is outlined in \cite{mcd17} and the monotonic spline approach is the method detailed in \cite{bonas21}. For the CMIP6-MIROC data, the mean coverage across all ensemble members as well as the standard deviation (in parenthesis) is reported.}
\label{tbl:UQComparison}
\end{centering}
\end{table}

\begin{figure}[!t]
\centering
\includegraphics[width = 10cm]{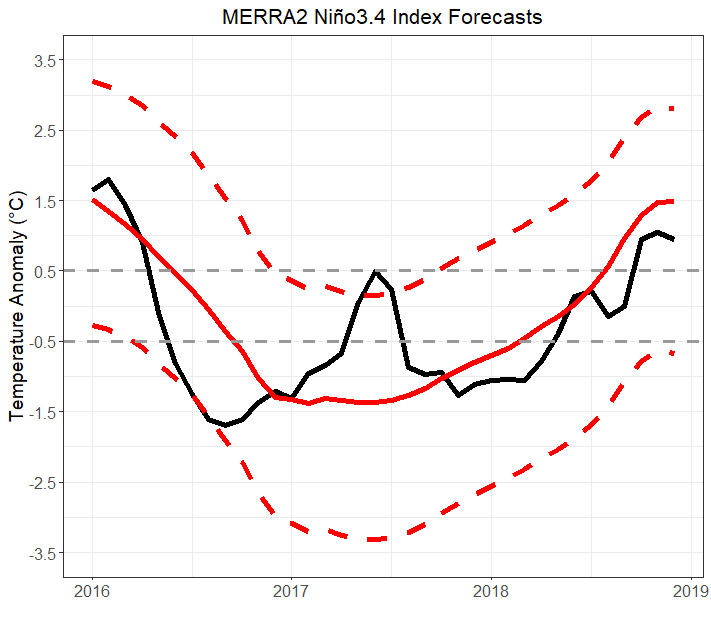}
\caption{Forecasts and uncertainty quantification for the Ni$\tilde{\text{n}}$o 3.4 index from the MERRA2 data with the DESN and the penalized quantile regression method.}
\label{fig:N34MERRA2Forcs}
\end{figure}

\begin{figure}[!bt]
\centering
\includegraphics[width = 13cm]{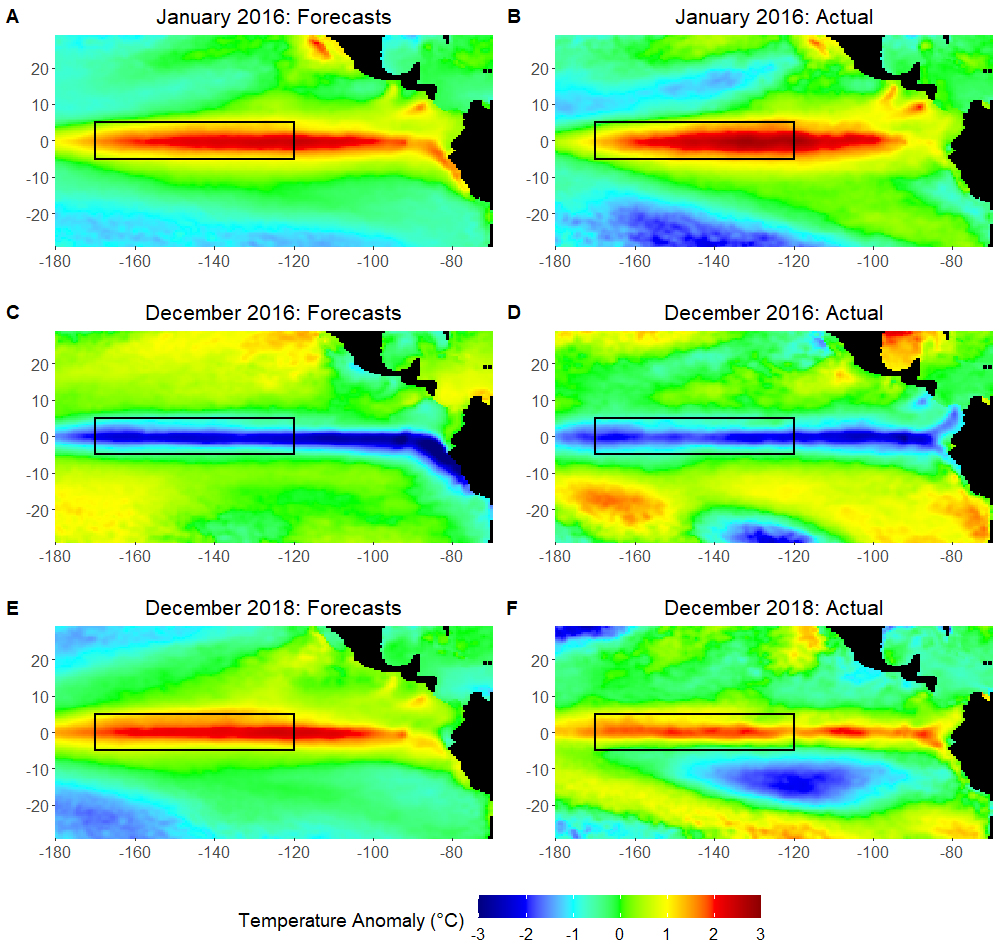}
\caption{For the MERRA2 data, forecasts at January 2016 ($K=1$, panel A), December 2016 ($K=12$, panel C) and December 2018 ($K=36$, panel E) with the corresponding true data (panels B, D, and F respectively) . Both January 2016 and December 2018 represent an El-Ni$\tilde{\text{n}}$o event and December 2016 represents a La-Ni$\tilde{\text{n}}$a event. The black box in all panels represents the Ni$\Tilde{\text{n}}$o-3.4 index region.}
\label{fig:MERRA2FullRegionForcs}
\end{figure}

Additionally, the calibration approach outlined in Section \ref{sec:Calibration} drastically improves the uncertainty quantification, has also shown in Table \ref{tbl:UQComparison}. Specifically, the nominal 95\% PIs generated from the ensemble-based approach in \cite{mcd17} only capture 61.1\% of the true data and the calibration method with monotonic splines in \cite{bonas21} improves the coverage only up to 77.8\%. The new approach outlined in Section \ref{sec:Calibration}, instead, increases the empirical coverage to 91.7\%, with an appreciable expansion of uncertainty through time. The median forecasts from the DESN model along with the corrected 95\% PIs can be seen in Figure \ref{fig:N34MERRA2Forcs}. Lastly, as was the case for the simulated data, it is of scientific interest and importance to produce forecasts over the entire spatial domain of interest. We recreated spatial grid created from the EOFs, and show how the model forecasts largely capture the ENSO magnitude across the entire domain in Figure \ref{fig:MERRA2FullRegionForcs}. It can be seen in this Figure how the DESN is able to accurately predict the SST anomalies across the whole spatial domain for multiple time steps ahead. In fact, it can be seen how the DESN effectively produces long-range forecasts as we show the predictions at $K=1$, $K=12$ and $K=36$ months ahead. These figures ultimately show how the proposed approach is able to produce accurate long-range forecasts over the entire spatial domain far into the future.

\section{Discussion}\label{sec:disc}

In this work we have introduced an approach at the interface between statistics and machine learning, a stochastic approximation of a dynamic neural network and we focused on one of the most established and challenging problems in environmental science: the prediction of quasi-periodic large-scale processes of the Earth systems whose occurrence ranges from a few years to multiple decades. While such processes are less understood than those dictated by the annual or daily cycle, their impacts on human habitability of the planet is far-reaching, especially in developing countries. 

On the methodological side, our work advocates the use of stochastic approximations of flexible yet highly parametric machine learning models in the context of dynamical multivariate (possibly spatial) data. This allows for fast linear algebra and hence fast predictions, a feature particularly desirable for high frequency sampling, where a prediction is only valuable to the extent that it can be produced at a fraction of the sampling rate. The computational convenience of a statistical model with machine learning dynamics comes at the cost of additional uncertainty in the forecast, but this can be controlled and calibrated with a \textit{post hoc} approach, as long as its computational cost is competitive with the sampling rate, and our proposed quantile regression method represents a convenient choice to model the uncertainty in long-range forecast while also allowing for some degree of robustness against the Gaussianity assumption. 

Principles and primitive equations such as conservation of physical quantities have been instrumental in the relentless surge of climate models and simulation-based climate science, and are now the standard of any impact assessment, including the periodic IPCC assessment reports, which have been the main reference for global actions on climate change coordinated by the United Nations. Despite the ever-increasing complexity of the physics-informed description of the Earth system, processes such as ENSO and PDO are still not fully characterized and have shown difficulty being adequately predicted. In this context, data-driven methods such as the one proposed in this work represent a powerful supplement to improve predictions, and models at the interface between statistics and machine learning represent an unprecedented opportunity to capture complex dynamics while also formally quantifying the uncertainty. Finally, hybrid physics-informed and data-driven methods do not necessarily have to use the former as a training set for the latter: recent approaches have also focused on how physical information can be effectively incorporated into a data-driven model \citep{kar21}. 

While dynamic neural networks have been originally devised as means to improve point prediction in a wide range of time series problems in computer science, we argue that a model-based stochastic approach, along the lines of Bayesian feed-forward neural networks \citep{mack92}, is a more appropriate and natural framework to cast these methods. To this end, while the proposed ESN was developed more than twenty years \citep{jae01}, its original intended purpose was to prevent numerical instabilities in gradient-based inference. A model-based perspective acknowledges that the improved stability of the DESN is not the ultimate goal, but is rather the consequence of a more parsimonious choice of parameters of a stochastic model, which also allows for a formal uncertainty quantification. Such perspective should, in the authors' view, be vigorously promoted by the statistical community in order to provide a long-lasting contribution in the debate on the role of artificial intelligence in the environmental sciences. 

While the proposed work focuses on large scale multi-year processes, the principle of augmenting environmental systems whose physical description is either incomplete or highly dependent on initial conditions with data-driven statistical methods is far broader. For example, turbulent processes at sub-kilometer scale have been shown to be even more challenging to predict with current physical modelling approaches such as large eddy simulations \citep{zhou14}, and data-driven models could provide and alternative and possibly more practical approach to forecasting turbulent properties. An improved knowledge of turbulent systems would then improve environmental applications ranging from urban air pollution to design and implementation of wind farms.

\bibliographystyle{chicago}
\bibliography{references}

\end{document}